\documentclass[twocolumn]{article}
\usepackage{graphicx}
\usepackage{cite}
\usepackage{epsfig}
\usepackage{fancyvrb}
\usepackage{amsmath,amssymb,amsfonts}
\usepackage{color}
\usepackage{times,mathtime}
\usepackage{sectsty}
\allsectionsfont{\mdseries}
\subsectionfont{\mdseries\itshape}
\subsubsectionfont{\mdseries\scshape}

\newcommand{\beq}{\begin{equation}}
\newcommand{\eeq}{\end{equation}}
\newcommand{\bea}{\begin{eqnarray}}
\newcommand{\eea}{\end{eqnarray}}
\newcommand   {\ev}[1]      {\langle #1\rangle}

\newcommand   {\fntt}       {FnIII${}_{10}$}
\newcommand   {\fntii}       {FnIII${}_{12}$}
\newcommand   {\fntiii}       {FnIII${}_{13}$}

\newcommand   {\DLcNI}      {\Delta L_\text{c}(\text{N}\to\text{I})}
\newcommand   {\avDLcNI}    {\overline{\Delta L_\text{c}(\text{N}\to\text{I})}}
\newcommand   {\DLcIU}      {\Delta L_\text{c}(\text{I}\to\text{U})}
\newcommand   {\DLcNU}      {\Delta L_\text{c}(\text{N}\to\text{U})}

\newcommand   {\Tw}         {T_\text{w}}
\newcommand   {\LI}         {L_\text{I}}
\newcommand   {\avLI}       {\bar L_\text{I}}
\newcommand   {\tI}         {\tau_\text{I}}

\newcommand   {\xu}         {x_\text{u}}
\newcommand   {\DLc}        {\Delta L_\text{c}}
\newcommand   {\Lc}         {L_\text{c}}

\newcommand   {\LN}         {L_\text{N}}
\newcommand   {\Ls}         {L_\text{s}}
\newcommand   {\Gs}         {G_\text{s}}
\newcommand   {\Gsc}        {G_\text{s}^\text{c}}
\newcommand   {\GN}         {G_\text{N}}
\newcommand   {\GNc}        {G_\text{N}^\text{c}}

\newcommand   {\FI}         {F_\text{I}}
\newcommand   {\avFI}       {\bar F_\text{I}}
\newcommand   {\Fc}         {F_\text{c}}
\newcommand   {\Eloc}       {E_\text{loc}}
\newcommand   {\Eev}        {E_\text{ev}}
\newcommand   {\Ehb}        {E_\text{hb}}
\newcommand   {\Ehp}        {E_\text{hp}}
\newcommand   {\Etot}       {E_\text{tot}}
\newcommand   {\kb}         {k_\text{B}}
\newcommand   {\nab}        {n_\text{AB}}
\newcommand   {\nbe}        {n_\text{BE}}
\newcommand   {\ncf}        {n_\text{CF}}
\newcommand   {\nfg}        {n_\text{FG}}

\newcommand  {\ARBB}     {{\it Annu.\ Rev.\ Bioph.\ Biom.\ }}

\newcommand  {\BJ}       {{\it Biophys.~J.\ }}
\newcommand  {\Cell}     {{\it Cell\ }}

\newcommand  {\EPL}      {{\it EPL\ }}
\newcommand  {\JBC}      {{\it J.~Biol.\ Chem.\ }}
\newcommand  {\JCC}      {{\it J.~Comput.\ Chem.\ }}
\newcommand  {\JCP}      {{\it J.~Chem.\ Phys.\ }}
\newcommand  {\JMB}      {{\it J.~Mol.\ Biol.\ }} 
\newcommand  {\JPCM}     {{\it J.~Phys.:\ Condens.\ Matter\ }}
\newcommand  {\JSM}      {{\it J.\ Stat.\ Mech.\ }}
\newcommand  {\Mac}      {{\it Macromolecules\ }} 
\newcommand  {\MB}       {{\it Matrix\ Biol.\ }} 
\newcommand  {\Nat}      {{\it Nature\ }} 
\newcommand  {\NRMCB}    {{\it Nat.\ Rev.\ Mol.\ Cell\ Biol.\ }}

\newcommand  {\PLOSB}    {{\it PLoS\ Biol.\ }}
\newcommand  {\PR}       {{\it Phys.\ Rev.\ }}
\newcommand  {\PRL}      {{\it Phys.\ Rev.\ Lett.\ }}
\newcommand  {\PNAS}     {{\it Proc.\ Natl.\ Acad.\ Sci.\ USA\ }}
\newcommand  {\Pro}      {{\it Proteins\ }}
\newcommand  {\ProSci}   {{\it Protein\ Sci.\ }}
\newcommand  {\Sci}      {{\it Science\ }}
\newcommand  {\Str}      {{\it Structure\ }}

\newcommand{\captionfonts}{\footnotesize}

\makeatletter  
\long\def\@makecaption#1#2{%
  \vskip\abovecaptionskip
  \sbox\@tempboxa{\captionfonts \textbf{#1}: #2}%
  \ifdim \wd\@tempboxa >\hsize
    {\captionfonts \textbf{#1}: #2\par}
  \else
    \hbox to\hsize{\hfil\box\@tempboxa\hfil}%
  \fi
  \vskip\belowcaptionskip}
\makeatother   

\begin{document}
\title{ Changing the mechanical unfolding pathway \\\vspace{0.2cm}
  of \fntt\ by tuning the pulling strength} 
\author{\normalsize Simon Mitternacht,${}^\ast$ Stefano Luccioli,${}^\dag$ 
  Alessandro Torcini,${}^\dag$ Alberto Imparato,${}^{\ddag\S}$ 
  and Anders Irb\"ack${}^\ast$}  
\date{\normalsize September 12, 2008}

\maketitle
\begin{small}

\noindent
\textit{We investigate the mechanical unfolding of the tenth type III
domain from fibronectin, \fntt, both at constant force and at constant
pulling velocity, by all-atom Monte Carlo simulations. We observe both
apparent two-state unfolding and several unfolding pathways involving
one of three major, mutually exclusive intermediate states. All the
three major intermediates lack two of seven native $\beta$-strands,
and share a quite similar extension.  The unfolding behavior is found
to depend strongly on the pulling conditions. In particular, we
observe large variations in the relative frequencies of occurrence for
the intermediates.  At low constant force or low constant velocity,
all the three major intermediates occur with a significant
frequency. At high constant force or high constant velocity, one of
them, with the N- and C-terminal $\beta$-strands detached, dominates
over the other two. Using the extended Jarzynski equality, we also
estimate the equilibrium free-energy landscape, calculated as a
function of chain extension. The application of a constant pulling
force leads to a free-energy profile with three major local
minima. Two of these correspond to the native and fully unfolded
states, respectively, whereas the third one can be associated with the
major unfolding intermediates.  }
\\\\
\rule[1pt]{50pt}{0.1mm}\\
\begin{footnotesize}
  ${}^\ast$Computational Biology \& Biological
  Physics, Department of Theoretical Physics, Lund University, Lund,
  Sweden. ${}^\dag$Istituto dei Sistemi Complessi, CNR, Sesto
  Fiorentino, Italy; INFN Sezione di Firenze, Sesto Fiorentino,
  Italy. ${}^\ddag$ISI Foundation, Torino, Italy.
  ${}^\S$Present address: Department of Physics and Astronomy, University of
  Aarhus, Denmark.\\\\
  Corresponding author: Simon Mitternacht, tel: +46-46-2223494,\\
  \hspace*{9pt}fax: +46-46-2229686, e-mail: simon@thep.lu.se \\\\
  Running title: Mechanical unfolding of \fntt \\\\ 
  Key words:
  force-induced unfolding, all-atom model, Monte Carlo, Jarzynski's
  equality.
\end{footnotesize}

\newpage

\noindent
Fibronectin is a giant multimodular protein that exists in both
soluble (dimeric) and fibrillar forms.  In its fibrillar form, it
plays a central role in cell adhesion to the extracellular matrix.
Increasing evidence indicates that mechanical forces exerted by cells
are a key player in initiation of fibronectin fibrillogenesis as well
as in modulation of cell-fibronectin adhesion, and thus may regulate
the form and function of fibronectin~\cite{Geiger:01,Vogel:06}.

Each fibronectin monomer contains more than 20 modules of three types,
called FnI--III. The most common type is FnIII, with $\sim$90 amino
acids and a $\beta$-sandwich fold. Two critical sites for the
interaction between cells and fibronectin are the RGD motif
Arg78-Gly79-Asp80~\cite{Ruoslahti:87} on the tenth FnIII module,
\fntt, and a synergistic site~\cite{Aoto:94} on the ninth FnIII
module, which bind to cell-surface integrins.  In the native structure
of \fntt, shown in Fig.~1, the RGD motif is found on the loop
connecting the C-terminal $\beta$-strands F and G. It has been
suggested that a stretching force can change the distance between
these two binding sites sufficiently to affect the cell-adhesion
properties, without deforming the sites
themselves~\cite{Vogel:06}. Force could also influence the adhesion
properties by causing full or partial unfolding of the \fntt\ module,
and thereby deformation of the RGD motif~\cite{Krammer:99}.  Whether
or not mechanical unfolding of fibronectin modules occurs in vivo is
controversial. It is known that cell-generated force can extend
fibronectin fibrils to several times their unstretched
length~\cite{Ohashi:99}. There are experiments indicating that this
extensibility is due to changes in quaternary structure rather than
unfolding~\cite{Abu-Lail:06}, while other experiments indicate that
the extensibility originates from force-induced unfolding of FnIII
modules~\cite{Baneyx:02,Smith:07}.  Also worth noting is that the
\fntt\ module is capable of fast refolding~\cite{Plaxco:97}.

\begin{figure}
  \begin{center}
    \includegraphics[width=0.45\textwidth]{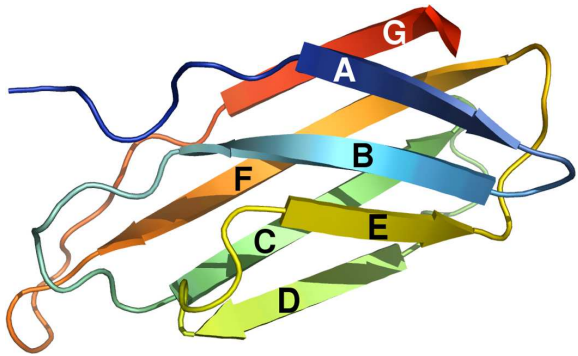}
  \end{center}
  
  \caption{Schematic illustration of the
    NMR-derived~\cite{Main:92} native structure of \fntt\ (Protein Data
    Bank ID 1ttf). Its seven $\beta$-strands are labeled A--G in sequence
    order.} 
\end{figure}

Atomic force microscopy (AFM) experiments have provided important
insights into the mechanical properties of FnIII
modules~\cite{Oberhauser:02,Li:05,Ng:07}.  Interestingly, it was found
that, although thermodynamically very stable~\cite{Cota:00}, the
cell-binding module \fntt\ is mechanically one of the least stable
FnIII modules~\cite{Oberhauser:02}. Further, it was shown that the
force-induced unfolding of \fntt\ often occurs through intermediate
states~\cite{Li:05}. While apparent one-step events were seen as well,
a majority of the unfolding events had a clear two-step
character~\cite{Li:05}.  A recent AFM study of pH
dependence~\cite{Ng:07} suggests that electrostatic contributions are
less important for the mechanical stability of \fntt\ than previously
thought.

Several groups have used computer simulations to investigate the
force-induced unfolding of
\fntt~\cite{Krammer:99,Paci:99,Klimov:00,Gao:02,Craig:04,Sulkowska:07,Li:07a}.
An early study predicted the occurrence of intermediate
states~\cite{Paci:99}. In these simulations, two unfolding pathways
were seen, both proceeding through partially unfolded intermediate
states. Both intermediates lacked two of the seven native
$\beta$-strands. The missing strands were A and B in one case, and A
and G in the other (for strand labels, see Fig.~1). A more recent
study reached somewhat different conclusions~\cite{Gao:02}. This study
found three different pathways, only one of which involved a partially
unfolded intermediate state, with strands A and B detached.  The
experiments~\cite{Li:05} are consistent with the existence of the two
different intermediates seen in the early simulations~\cite{Paci:99},
but do not permit an unambiguous identification of the states. When
comparing the experiments with these simulations, it should be kept in
mind that the forces studied in the simulations were larger than those
studied experimentally.

Here we use an implicit-water all-atom model with a simple and
computationally convenient energy function~\cite{Irback:03,Irback:05a}
to investigate how the response of \fntt\ to a stretching force
depends on the pulling strength. We study the unfolding
behavior both at constant force and at constant pulling velocity.  
Some previous studies were carried out using 
explicit-solvent models~\cite{Krammer:99,Gao:02,Craig:04}. These 
models might capture important details that our implicit-solvent model 
ignores, like weakening of specific hydrogen bonds through interactions
between water molecules and the protein backbone~\cite{Lu:00}.  
The advantage of our model is computational convenience. 
The relative simplicity of the model makes it possible for us to
generate a large set of unfolding events, which is important
when studying a system with multiple unfolding pathways.

Our analysis of the generated unfolding trajectories consists of two
parts. The first part aims at characterizing the major unfolding
pathways and unfolding intermediates.  In the second part, we use the
extended Jarzynski equality
(EJE)~\cite{Jarzynski:97,Crooks:99,Hummer:01} to estimate the
equilibrium free-energy landscape, calculated as a function of end-to-end
distance. This analysis extends previous work on simplified protein
models~\cite{West:06a,Imparato:07a,Imparato:07c,Imparato:07b} to an
atomic-level model. This level of detail may be needed to facilitate
comparisons with future EJE reconstructions based on experimental
data. Indeed, two applications of this method to experimental protein data
were recently reported~\cite{Harris:07, Imparato:08a}.

\section*{Model and Methods} 

\subsection*{Model}

We use an all-atom model with implicit water,
torsional degrees of freedom,   
and a simplified energy function~\cite{Irback:03,Irback:05a}.   
The energy function
\begin{equation}
  E=\Eloc+\Eev+\Ehb+\Ehp
\label{eo}
\end{equation}
is composed of four terms.  The term $\Eloc$ is local in sequence and
represents an electrostatic interaction between adjacent peptide units
along the chain. The other three terms are non-local in sequence.  The
excluded volume term $\Eev$ is a $1/r^{12}$ repulsion between pairs of
atoms.  $E_{\mathrm{hb}}$ represents two kinds of hydrogen bonds:
backbone-backbone bonds and bonds between charged side chains and the
backbone. The last term $\Ehp$ represents an effective hydrophobic
attraction between nonpolar side chains.  It is a simple pairwise
additive potential based on the degree of contact between two nonpolar
side chains. The precise form of the different interaction terms and
the numerical values of all geometry parameters can be found
elsewhere~\cite{Irback:03,Irback:05a}.

It has been shown that this model, despite its simplicity, provides a
good description of the structure and folding thermodynamics of
several peptides with different native geometries~\cite{Irback:05a}.
For the significantly larger protein \fntt, it is computationally
infeasible to verify that the native structure is the global
free-energy minimum.  However, in order to study unfolding, it is
sufficient that the native state is a local free-energy minimum. In
our model, with unchanged parameters~\cite{Irback:03,Irback:05a}, 
the native state of \fntt, indeed, is a long-lived state 
corresponding to a free-energy minimum, as will be seen below. 

The same model has previously been used to 
study both mechanical and thermal unfolding of
ubiquitin~\cite{Irback:05b,Irback:06a}. In agreement 
with AFM experiments~\cite{Schlierf:04}, it was  
found that ubiquitin, like \fntt, displays a mechanical unfolding 
intermediate far from the native state, and this intermediate
was characterized~\cite{Irback:05b}. The picture emerging from 
this study~\cite{Irback:05b} was subsequently supported by  
ubiquitin simulations based on completely different 
models~\cite{Li:07b,Kleiner:07,Imparato:08b}.  

The energy function $E$ of Eq.~\ref{eo} describes an unstretched
protein. In our calculations, the protein is pulled either by a
constant force or with a constant velocity. In the first case,
constant forces $-\vec{F}$ and $\vec{F}$ act on the N and C termini,
respectively. The full energy function is then given by
\begin{equation}
  \Etot= E- \vec{F} \cdot \vec{R}
\end{equation}
where $\vec{R}$ is the vector from the N to the C terminus.  In the
constant-velocity simulations, the pulling of the protein is modeled
using a harmonic potential in the end-to-end distance $L=|\vec R|$
whose minimum $L_v(t)$ varies linearly with Monte Carlo (MC) time
$t$. With this external potential, the full, time-dependent energy
function becomes
\begin{equation}
  \Etot(t) = E + \frac k2 \left[L_v(t) - L\right]^2= 
              E+\frac k2 \left[L_0+vt-L\right]^2 
\label{ev}\end{equation}
where $k$ is a spring constant, $v$ is the pulling velocity, and $L_0$
is the initial equilibrium position of the spring. The spring
constant, corresponding to the cantilever stiffness in AFM experiments, 
is set to $k=37$\,pN/nm. The experimental \fntt\ study of~\cite{Li:05}   
reported a typical spring constant of $k\sim50$\,pN/nm.

\subsection*{Simulation methods}

Using MC dynamics, we study six constant force magnitudes $F$ (50\,pN,
80\,pN, 100\,pN, 120\,pN, 150\,pN and 192\,pN) and four constant
pulling velocities $v$ (0.03\,fm/MC step, 0.05\,fm/ MC step,
0.10\,fm/MC step and 1.0\,fm/MC step), at a temperature of
288~K. Three different types of MC updates are used: (i) Biased
Gaussian Steps~\cite{Favrin:01}, BGS, which are semi-local updates of
backbone angles; (ii) single-variable Metropolis updates of side-chain
angles; and (iii) small rigid-body rotations of the whole chain.  The
BGS move simultaneously updates up to eight consecutive backbone
angles, in a manner that keeps the chain ends approximately fixed. In
the constant-velocity simulations, the time-dependent parameter
$L_v(t)$ is changed after every attempted MC step.

As a starting point for our simulations, we use a model approximation
of the experimental \fntt\ structure (backbone root-mean-square
deviation $\approx$\,0.2\,nm), obtained by simulated annealing. All
simulations are started from this initial structure, with different
random number seeds. However, in the constant-velocity runs, the
system is first thermalized in the potential $E+k(L_0-L)^2/2$ for
$10^7$ MC steps ($L_0=3.8$\,nm), before the actual simulation is
started at $t=0$. The thermalization is a prerequisite for the
Jarzynski analysis (see below).

The constant-force simulations are run for a fixed time, which depends
on the force magnitude. There are runs in which the protein remains
folded over the whole time interval studied.  The constant-velocity
simulations are run until the spring has been pulled a distance of
$vt= 35$\,nm. At this point, the protein is always unfolded.

Our simulations are carried out using the program package
PROFASI~\cite{Irback:06b}, which is a C++ implementation of this
model. 3D structures are drawn with PyMOL~\cite{pymol}.

\subsection*{Analysis of pathways and intermediates}

To characterize pathways and intermediates, we study the evolution of
the native secondary-structure elements along the unfolding
trajectories. For this purpose, during the course of the simulations,
all native hydrogen bonds connecting two $\beta$-strands (see Fig.~1)
are monitored. A bond is defined as present if the energy of that bond
is lower than a cutoff ($-2.4\kb T$). Using this data, we can describe
a configuration by which pairs of $\beta$-strands are formed. A
$\beta$-strand pair is said to be formed if more than a fraction 0.3
of its native hydrogen bonds are present. Whether individual
$\beta$-strands are present or absent is determined based on which
$\beta$-strand pairs the conformation contains.

The characterization of intermediate states requires slight\-ly
different procedures in the respective cases of constant force and
constant velocity.  For constant force, a histogram of the end-to-end
distance $L$, covering the interval
$3\,\textrm{nm}<L<27\,\textrm{nm}$, is made for each unfolding
trajectory.  Each peak in the histogram corresponds to a metastable
state along the unfolding pathway. To reduce noise the histogram is
smoothed with a sliding $L$ window of 0.3\,nm. Peaks higher than a
given cutoff are identified. Two peaks that are close to each other
are only considered separate states if the values between them drop
below half the height of the smallest peak. The position of an
intermediate, $\LI$, is calculated as a weighted mean over the
corresponding peak. The area under the peak provides, in principle, a
measure, $\tI$, of the life time of the state. However, due to
statistical difficulties, we do not measure average life times of
intermediate states.
          
In the constant-velocity runs, the unraveling of the native state or
an intermediate state is associated with a rupture event, at which a
large drop in force occurs.  To ascertain that we register actual
rupture events and not fluctuations due to thermal noise, the force
versus time curves are smoothed with a sliding time window of
$\Tw=0.3\,\textrm{nm}/v$, where $v$ is the pulling velocity.  Rupture
events are identified as drops in force that are larger than 25\,pN
within a time less than $\Tw$. The point of highest force just before
the drop defines the rupture force, $\FI$, and the 
end-to-end distance, 
$\LI$, of the corresponding state.  Only rupture events with a time
separation of at least $2\Tw$ are considered separate events.  The
rupture force $\FI$ is a stability measure statistically 
easier to estimate than the
life time $\tI$ at constant force.
  
For a peak with a given $\LI$, to decide which $\beta$-strands the
corresponding state contains, we consider all stored configurations
with $|L-\LI|<0.1$\,nm. All $\beta$-strand pairs occurring at least
once in these configurations are considered formed in the state. With
this prescription, it happens that separate peaks from a single run
exhibit the same set of $\beta$-strand pairs. Distinguishing between
different substates with the same secondary-structure elements is
beyond the scope of the present work. Such peaks are counted as a
single state, with $\LI$ set to the weighted average position of the
merged peaks.
  
\subsection*{Jarzynski analysis}

From the constant-velocity trajectories, we estimate the 
equilibrium free-energy
landscape $G_0(L)$, as a function of the end-to-end distance $L$, for
the unstretched protein by using
EJE~\cite{Jarzynski:97,Crooks:99,Hummer:01,Imparato:06}. For our
system, this identity takes the form
\begin{multline}
  \textrm{e}^{-G_0(L)/\kb T} =
  \textrm{constant}\cdot\textrm{e}^{k[L-L_v(t)]^2/2\kb T} \times \\
  \ev{\delta(L-L(C_t))\textrm{e}^{-W_t/\kb T}}_t
  \label{ej}
\end{multline}
where $\kb$ is Boltzmann's constant, $T$ the temperature, and $C_t$
stands for the configuration of the system at time $t$. In this
equation, $\ev{\ldots}_t$ denotes an average over trajectories
$C_\tau$, $0<\tau<t$, with starting points $C_0$ drawn from the
Boltzmann distribution corresponding to $\Etot(0)$ (see
Eq.~\ref{ev}). The quantity $W_t$ is the work done on the system along
a trajectory and is given by
\begin{equation}
  W_t=\int_0^t kv[L_v(\tau)-L(C_\tau)]\,\text{d}\tau=\int F\, \text{d} L_v
\end{equation}

As discussed in~\cite{Hummer:01,Imparato:06}, combining Eq.~\ref{ej}
with the weighted histogram method~\cite{Ferrenberg:89}, one finds
that the optimal estimate of the target function $G_0(L)$ is given by
\begin{multline}
  G_0(L)= \\-\kb T\ln
  \left[\frac{\sum_t\ev{\delta(L-L(C_t))\textrm{e}^{-W_t/\kb T}}_t/
      \ev{\textrm{e}^{-W_t/\kb T}}}
    {\sum_t\textrm{e}^{-k[L-L_v(t)]^2/2\kb T}/\ev{\textrm{e}^{-W_t/\kb T}}}\right],
  \label{eqhis}
\end{multline} 
up to an additive constant. As in an experimental situation, for each
unfolding trajectory, we sample the 
end-to-end distance $L(C_t)$ and the work
$W_t$ at discrete time intervals $k \Delta \tau$, with $k=0,\dots,n$
and $n\Delta \tau=t$. The sums appearing in Eq.~\ref{eqhis} thus run
over these discrete times.

Let $L_{\min}$ and $L_{\max}$ be the minimal and maximal end-to-end distances,
respectively, observed in the unfolding trajectories. We divide the
interval $\left[L_{\min},L_{\max}\right]$ into sub-intervals of length
$\Delta L$ and evaluate $G_0(L_i)$ for each $L_i=L_{\min}+(i+1/2)
\Delta L$ by exploiting Eq.~\ref{eqhis}. The two averages appearing in
this equation are estimated as
$\overline{\theta_i(L(C_t))\exp(-W_t/\kb T)}$ and
$\overline{\exp(-W_t/\kb T)}$, where the bar indicates an average over
trajectories and the function $\theta_i(x)$ is defined as
$\theta_i(x)=1$ if $|x-L_i|<\Delta L/2$ and $\theta_i(x)=0$
otherwise. Further details on the scheme used can be found in
\cite{Imparato:06}.

\section*{Results}

\subsection*{Description of  the calculated unfolding traces}

We study the mechanical unfolding of \fntt\ for six constant forces
and four constant velocities. Table~1 shows the number of runs and the
length of each run in these ten cases. At low force or low velocity,
it takes longer for the protein to unfold, which makes it necessary to
use longer and computationally more expensive trajectories.

\begin{table}
  \begin{footnotesize}
    \begin{center}
    \begin{tabular} {rrrr}
      pulling force or velocity& runs & MC steps/$10^6$\\
      \hline\hline
      50\,pN  &  98 & 1\,000\\
      80\,pN  & 100 & 1\,000\\
      100\,pN & 100 & 250\\
      120\,pN & 200 & 100\\
      150\,pN & 340 & 50\\
      192\,pN & 600 & 30\\
      \hline
      0.03\,fm$/$MC step & 100 & 1\,167\\
      0.05\,fm$/$MC step & 99 & 700\\
      0.10\,fm$/$MC step & 99 & 350\\
      1.0\,fm$/$MC step & 200 & 35\\
      \hline\hline
    \end{tabular}
    \end{center}
  \end{footnotesize}  
  \caption{Number of runs and the length of each run, in
    number of elementary MC steps, at the different pulling conditions
    studied.}    
\end{table}

Fig.~2 shows the time evolution of the end-to-end distance $L$ in a
representative set of runs at constant force (100\,pN). Typically each
trajectory starts with a long waiting phase with $L\sim5$\,nm, where
the molecule stays close to the native conformation.  In this phase,
the relative orientation of the two $\beta$-sheets (see Fig.1) might
change, but all native $\beta$-strands remain unbroken. 
The waiting phase is followed by a sudden increase in $L$. This step
typically leads either directly to the completely unfolded state with
$L\sim 30$\,nm or, more commonly, to an intermediate state at $L\sim
12$--16\,nm. The intermediate is in turn unfolded in another abrupt
step that leads to the completely stretched state. In a small fraction
of the trajectories, depending on force, the protein is still in the
native state or an intermediate state when the simulation
stops. Intermediates outside the range 12--16\,nm are unusual but
occur in some runs. For example, a relatively long-lived intermediate
at 21\,nm can be seen in one of the runs in Fig.~2.

\begin{figure}
  \begin{center}
    \includegraphics[width=0.47\textwidth]{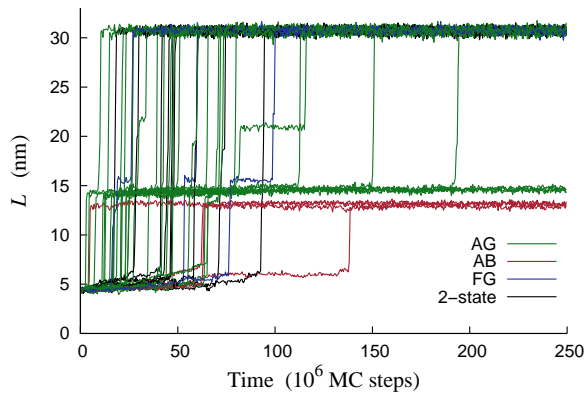}
  \end{center}

  \caption{MC time evolution of the end-to-end distance in 42
    independent simulations with a constant pulling force of 100\,pN.
    The three most frequent intermediates lack different pairs of
    native $\beta$-strands: AG, FG, or AB. Trajectories in which these
    states occur are labeled green, blue and red,
    respectively. Apparent two-state events are colored black.}
\end{figure}

Fig.~3 shows samples of unfolding traces at constant velocity
(0.05\,fm/MC step). Here force is plotted against end-to-end distance.
As in the constant-force runs, there are two main events in most
trajectories. First, the native state is pulled until it ruptures at
$L\sim5$\,nm.  The chain is then elongated without much resistance
until it, in most cases, reaches an intermediate at
$L\sim12$--16\,nm. Here the force increases until there is a second
rupture event. After that, the molecule is free to elongate towards
the fully unfolded state with $L\sim 30$\,nm. Some trajectories have
force peaks at other $L$.  An unusually large peak of this kind can be
seen at 22\,nm in Fig.~3. Inspection of the corresponding structure
reveals that it contains a three-stranded $\beta$-sheet composed of
the native CD hairpin and a non-native strand. This sheet is pulled
longitudinally, which explains why the stability is high.  Another
feature worth noting in Fig.~3 is that the pulling velocity is
sufficiently small to permit the force to drop to small values between
the peaks.
\begin{figure}
  \begin{center}
    \includegraphics[width=0.47\textwidth]{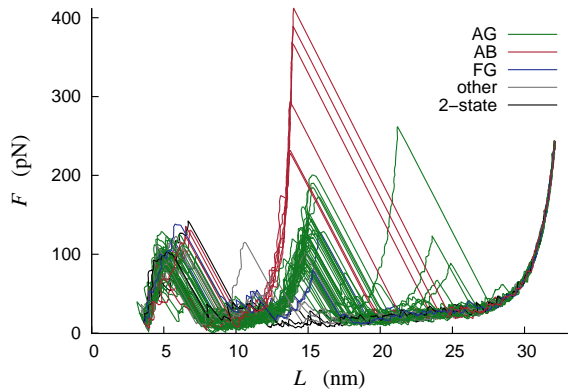}
  \end{center}
  \caption{Force versus end-to-end distance in 55 independent
    simulations with a constant pulling velocity of
    $0.05$\,fm/MC~step. Noise has been filtered out using a sliding
    time window of $6\cdot10^6$\,MC steps.  The color coding is the
    same as in Fig.~2, with the addition of a new category for a few
    trajectories not belonging to any of the four categories in that
    figure. These trajectories are colored grey.}
\end{figure}

There are several similarities between the unfolding events seen at
constant force and at constant velocity. In most trajectories, there are
stable intermediates, and the unfolding from both the native and
intermediate states is abrupt. Also, the vast majority of the observed
intermediates have a similar end-to-end distance, in the range
12--16\,nm. It should be noticed that experiments typically
measure contour-length differences rather than end-to-end distances.
Below we analyze contour-length differences between the native state and
our calculated intermediates, which turn out to be in good agreement with 
experimental data.

The trajectories can be divided into three categories: apparent
two-state unfolding, unfolding through intermediate states, and
trajectories in which no unfolding takes place. Table~2 shows the
relative frequencies of these groups at the different pulling
conditions. The number of trajectories in which the protein remains
folded throughout the run obviously depends on the trajectory
length. More interesting to analyze is the ratio between the two kinds
of unfolding, with or without intermediate states. In the
constant-force runs, this ratio depends strongly on the magnitude of
the applied force; unfolding through intermediates dominates at the
lowest force, but is less common than apparent two-state unfolding at
the highest force. In the constant-velocity runs, unfolding through
intermediates is much more probable than apparent two-state unfolding
at all the velocities studied.
\begin{table}
  \begin{center}
  \begin{footnotesize}
    \begin{tabular} {rrrr}
      pulling force or velocity & $n=2$  & $n\geq 3$ & no unfolding \\
      \hline\hline
               50\,pN  & 0.01 & 0.79 & 0.20 \\
	       80\,pN  & 0.21 & 0.79 & 0 \\
               100\,pN & 0.23 & 0.77 & 0 \\
	       120\,pN & 0.24 & 0.76 & 0 \\
	       150\,pN & 0.29 & 0.72 & $<$0.01 \\
	       192\,pN & 0.54 & 0.46 & 0\\\hline
      0.03\,fm/MC step & 0.04 & 0.96 & 0\\
      0.05\,fm/MC step & 0.07 & 0.93 & 0\\
      0.10\,fm/MC step & 0.03 & 0.97 & 0\\
       1.0\,fm/MC step &    0 &  1.0 & 0\\
      \hline\hline
    \end{tabular}
  \end{footnotesize}
  \end{center}
  \caption{The fractions of trajectories in which unfolding occurs
    either in an apparent two-state manner (labeled $n=2$) or through
    intermediate states (labeled $n\geq3$). ``No unfolding'' refers to
    the fraction of trajectories in which the protein remains folded
    throughout the run (with $L<8$\,nm).}
\end{table}
 
\subsection*{Identifying pathways and intermediates}

The fact that most observed intermediates fall in the relatively
narrow $L$ interval of 12--16\,nm does not mean that they are
structurally similar. Actually, the data in Figs.~2 and 3 clearly
indicate that these intermediates can be divided into three groups
with similar but not identical end-to-end distances. The
$\beta$-strand analysis (see Model and Methods) reveals that these
three groups correspond to the detachment of different pairs of
$\beta$-strands, namely A and G, A and B, or F and G.  The prevalence
of these particular intermediate states is not surprising, given the
native topology.  When pulling the native structure of \fntt, the
interior of the molecule is shielded from force by the N- and
C-terminal $\beta$-strands, A and G. Consequently, in 95\,\% or more
of our runs, either strand A or G is the first to detach, for all the
pulling conditions studied.  Most commonly, this detachment is
followed by a release of the other strand of the two. But, when A (G)
is detached, B (F) is also exposed to force. We thus have three main
options for detaching two strands, AG, AB or FG, which actually
correspond to the three major intermediates we observe.

Intermediates outside the interval 12--16\,nm also occur in our
simulations.  When applied to the intermediates with $L<12$\,nm, the
$\beta$-strand analysis identifies two states with one strand
detached, A or G. The intermediates with $L>16$\,nm are scattered in
$L$ and correspond to rare states with more than two strands
detached. The intermediate at 21\,nm seen in one of the runs in Fig.~2
lacks, for example, four strands (A, B, F and G).  However, in these
relatively unstructured states with more than two strands detached,
the remaining strands are often disrupted, which makes the binary
classification of strands as either present or absent somewhat
ambiguous. Moreover, it is not uncommon that these large-$L$
intermediates contain some non-native secondary structure.  In what
follows, we therefore focus on the five states seen with only one or
two strands detached.

For convenience, the intermediates will be referred to by which
strands are detached. The intermediate with strands A and B unfolded
will thus be labeled AB, etc.  Tables~3 and 4 show basic properties of
the A, G, AB, AG and FG intermediates, as observed at constant force
and constant velocity, respectively.

\begin{table*}
  \begin{center}
  \begin{footnotesize}
    \begin{tabular} {rrrrrrrrrrrrr}
      \multicolumn{1}{r}{}
             & \multicolumn{2}{c}{ 50\,pN}
             & \multicolumn{2}{c}{ 80\,pN}
             & \multicolumn{2}{c}{100\,pN}
             & \multicolumn{2}{c}{120\,pN}
             & \multicolumn{2}{c}{150\,pN}
             & \multicolumn{2}{c}{192\,pN} \\
      \hline\hline
      state
            & $f$ & $\avLI$
            & $f$ & $\avLI$
            & $f$ & $\avLI$
            & $f$ & $\avLI$
            & $f$ & $\avLI$
            & $f$ & $\avLI$ \\
      \hline
         AG & 0.46 & 13.9 & 0.49 & 14.3 & 0.65 & 14.3 & 0.69 & 14.5 &    0.69 & 14.6 &    0.45 & 14.7\\
         AB & 0.35 & 12.4 & 0.14 & 12.9 & 0.09 & 13.1 & 0.03 & 13.2 & $<$0.01 & ---  & $<$0.01 & --- \\
         FG & 0.15 & 14.8 & 0.13 & 15.2 & 0.03 & 15.5 & 0.03 & 15.7 & $<$0.01 & ---  & $<$0.01 & --- \\
          G & 0.19 & 11.1 & 0.04 & 11.8 &    0 & ---  & 0    & ---  &       0 & ---  &       0 & --- \\
          A & 0.13 &  6.7 &    0 &  --- &    0 & ---  & 0    & ---  &       0 & ---  &       0 & --- \\
      \hline\hline
    \end{tabular}
  \end{footnotesize}
  \end{center}
  \caption{Frequency $f$ and average extension $\avLI$ (in nm) of
    intermediate states in the constant-force simulations.  The label
    of a state indicates which $\beta$-strands are detached, that is
    the state AG lacks strands A and G, etc. The frequency $f$ is the
    number of runs in which a given state was seen, divided by the
    total number of runs in which unfolding occurred. The statistical
    uncertainties on $\avLI$ are about 0.1\,nm or smaller. ``---''
    indicates not applicable. }
\end{table*}

\begin{table*}
  \begin{center}
  \begin{footnotesize}
    \begin{tabular} {rrrrrrrrrrrrr}
      \multicolumn{1}{r}{} &
      \multicolumn{3}{c}{0.03 fm$/$MC step}  &
      \multicolumn{3}{c}{0.05 fm$/$MC step}  &
      \multicolumn{3}{c}{0.10 fm$/$MC step}  &
      \multicolumn{3}{c}{1.0 fm$/$MC step} \\
      \hline\hline
      state
      & $f$ & $\avFI$ & $\avLI$
      & $f$ & $\avFI$ & $\avLI$
      & $f$ & $\avFI$ & $\avLI$
      & $f$ & $\avFI$ & $\avLI$\\
      \hline
         AG & 0.60 & 115 & 14.9 & 0.69 & 121 & 14.9 & 0.78 & 131 & 14.8 & 0.81 & 198 & 15.0 \\
         AB & 0.14 & 283 & 13.7 & 0.09 & 289 & 13.8 & 0.08 & 333 & 13.9 & 0.04 & 318 & 13.9 \\
         FG & 0.15 & 119 & 15.6 & 0.08 & 107 & 15.3 & 0.08 & 162 & 16.0 & 0.04 & 216 & 15.7 \\
          G & 0.05 &  54 & 10.5 & 0.08 &  73 & 10.8 & 0.20 &  46 &  9.9 & 0.06 &  67 & 10.3 \\
          A & 0.06 &  43 &  6.2 & 0.07 &  53 &  7.2 & 0.09 &  57 &  6.9 & 0.03 &  81 &  7.2 \\
      \hline\hline
    \end{tabular}
  \end{footnotesize}
  \end{center}
  \caption{Frequency $f$, average rupture force $\avFI$ (in pN) and
    average extension $\avLI$ (in nm) of intermediate states in the
    constant-velocity simulations. The statistical uncertainties are
    10--20\,\% on $\FI$, about 0.1\,nm or smaller on $\LI$ for AG and
    AB, and about 0.5\,nm on $\LI$ for FG, G and A.  }
\end{table*}

From Tables~3 and 4, several observations can be made.  A first one is
that the average end-to-end distance, $\avLI$, of a given state
increases slightly with increasing force.  More importantly, it can be
seen that the relative frequencies with which the different
intermediates occur depend strongly on the pulling conditions. At high
force or high velocity, the AG intermediate stands out as the by far
most common one. By contrast, at low force or low velocity, there is
no single dominant state. In fact, at $F=50$\,pN as well as at
$v=0.03$\,fm/MC step, all the five states occur with a significant
frequency.

Table~4 also shows the average rupture force, $\avFI$, of the
different states, at the different pulling velocities.  Although the
data are somewhat noisy, there is a clear tendency that $\avFI$, for a
given state, slowly increases with increasing pulling velocity, which
is in line with the expected logarithmic $v$
dependence~\cite{Evans:97}.  Comparing the different states, we find
that those with only one strand detached (A and G) are markedly weaker
than those with two strands detached (AG, AB and FG), as will be
further discussed below. Most force-resistant is the AB
intermediate. This state occurs much less frequently than the AG
intermediate, especially at high velocity, but is harder to break once
formed.  Compared to experimental data, our $\avFI$ values for the
intermediates are somewhat large. The experiments found a relatively
wide distribution of unfolding forces centered at
40--50\,pN~\cite{Li:05}, which is a factor two or more lower than what
we find for the AG, AB and FG intermediates. Our results for the
unfolding force of the native state are consistent with experimental
data.  For the native state, the experiments found unfolding forces of
$75\pm20$\,pN~\cite{Oberhauser:02} and $90\pm20$\,pN~\cite{Li:05}.
Our corresponding results are $88\pm2$\,pN, $99\pm2$\,pN and
$114\pm3$\,pN at $v=0.03$\,fm/MC step, $v=0.05$\,fm/MC step and
$v=0.10$\,fm/MC step, respectively.

The AG, AB and FG intermediates do not only require a significant rupture
force in our constant-velocity runs, but are also long-lived in our
constant-force simulations. In fact, in many runs, the system is still
in one of these states when the simulation ends, which means that
their average life times, unfortunately, are too long to be determined
from the present set of simulations. Nevertheless, there is a clear trend 
that the AB intermediate is more long-lived than the other two, which in 
turn have similar life times. The relative life times of these states in
the constant-force runs are thus fully consistent with their 
force-resistance in the constant-velocity runs.

At high constant force, we see a single dominant intermediate, the AG
state, but also a large fraction of events without any detectable
intermediate. Interestingly, it turns out that the same two strands, A
and G, are almost always the first to break in the apparent two-state
events as well. Table~5 shows the fraction of all trajectories, with
or without intermediates, in which A and G are the first two strands
to break, at the different forces studied. At 192\,pN, this fraction
is as large as 98\,\%. Although the time spent in the state with
strands A and G detached varies from run to run, there is thus an
essentially deterministic component in the simulated events at high
force.

\begin{table}
  \begin{center}
  \begin{footnotesize}
    \begin{tabular} {rrrrrrr}
      \multicolumn{1}{r}{\!first pair} &
      \multicolumn{1}{r}{50\,pN}   &
      \multicolumn{1}{r}{80\,pN}   &
      \multicolumn{1}{r}{100\,pN}  &
      \multicolumn{1}{r}{120\,pN}  &
      \multicolumn{1}{r}{150\,pN}  &
      \multicolumn{1}{r}{192\,pN}  \\
      \hline\hline
         A \& G & 0.50 &  0.69 & 0.87 &  0.935 & 0.973 & 0.980\\
         A \& B & 0.35 &  0.15 & 0.09 &  0.025 & 0.006 & 0.007\\ 
         F \& G & 0.15 &  0.16 & 0.04 &  0.040 & 0.021 & 0.013\\ 
      \hline\hline
    \end{tabular}
  \end{footnotesize}
  \end{center}
  \caption{The fractions of all unfolding events in which 
    the first two strands to break are A \& G, F \& G, and A \& B, respectively, 
    at different constant forces. The first pair to break was always one of
    these three.}
\end{table}

The unfolding behavior at low force or velocity is, by contrast,
complex, with several possible pathways. Fig.~4 illustrates the
relations between observed pathways at the lowest pulling velocity,
0.03\,fm/MC step. The main unfolding path begins with the detachment
of strand G, followed by the formation of the AG intermediate, through
the detachment of A.  There are also runs in which the same
intermediate occurs but A and G detach in the opposite order.  Note
that for the majority of the trajectories the boxes A and G in Fig.~4
only indicate passage through these states, not the formation of an
intermediate state. In a few events, it is impossible to say which
strand breaks first. In these events, the initial step is either that
the hairpin AB detaches as one unit, or that strands A and G are
unzipped simultaneously. Detachment of the FG hairpin in one chunk
does not occur in the set of trajectories analyzed for
Fig.~4. Finally, we note that in the few trajectories where G occurs
as an intermediate, the FG intermediate is always visited as well, but
never AG.  Similarly, the few trajectories where the A intermediate
occurs also contain the AG intermediate, but not AB.  We find no
example where the AB intermediate is preceded by another intermediate.

\begin{figure*}
  \begin{center}
    \includegraphics[width=12cm]{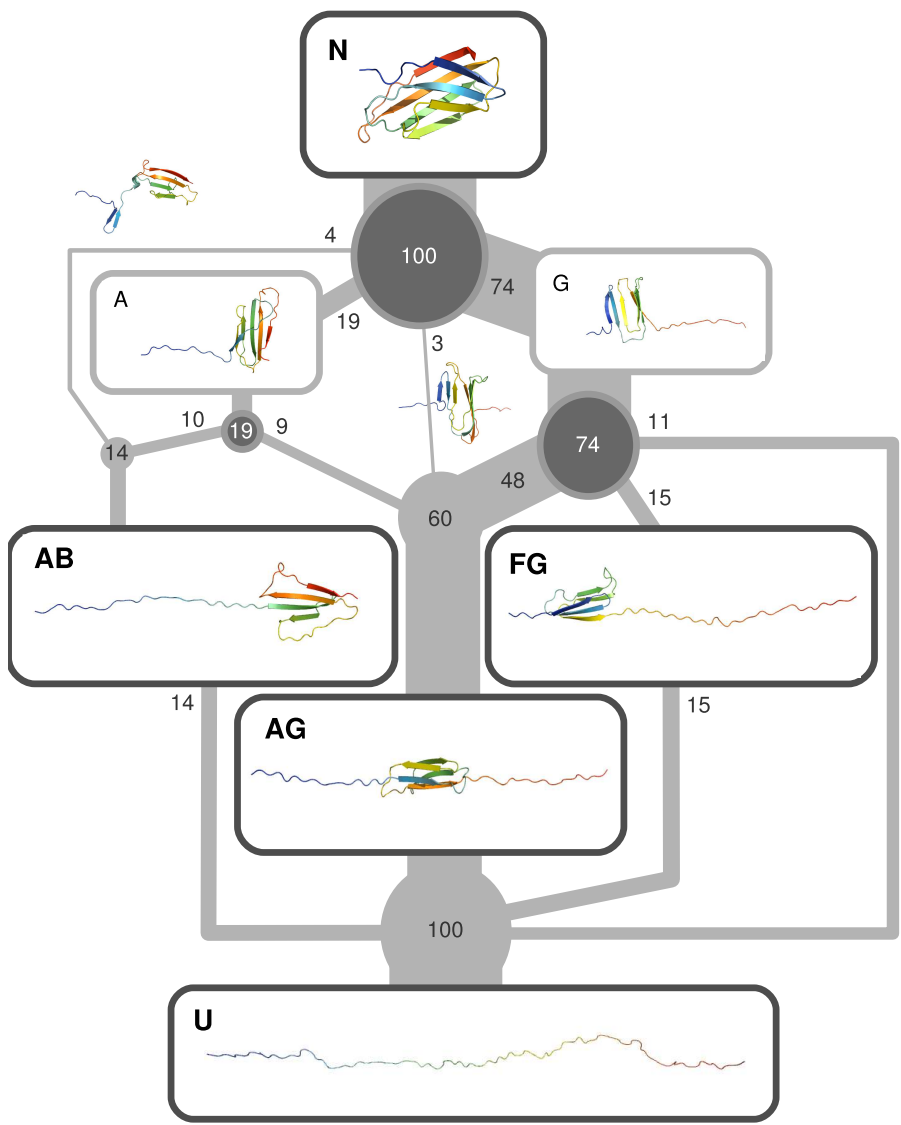}
  \end{center}
  \caption{Illustration of the diversity of unfolding pathways in the
    100 constant-velocity unfolding simulations at
    $v=0.03$\,fm/MC~step. The numbers indicate how many of the
    trajectories follow a certain path. The boxes illustrate important
    structures along the pathways and boxes with dark rims correspond
    to the most long-lived states. Dark circles mark branch
    points. Most trajectories pass through G or A, but only a fraction
    spend a significant amount of time there (see Table 4). The line
    directly from G to U corresponds to events that either have no
    intermediate at all or only have intermediates other than the main
    three. The direct lines N$\rightarrow$AB and N$\rightarrow$AG
    describe events that do not clearly pass through A or G and
    examples of structures seen in those events are illustrated by the
    unboxed cartoons next to the lines.}
\end{figure*}

The unfolding pattern illustrated in Fig.~4 can be partly understood
by counting native hydrogen bonds. The numbers of hydrogen bonds
connecting the strand pairs AB, BE, CF and FG are $\nab=7$, $\nbe=5$,
$\ncf=8$ and $\nfg=6$, respectively.  In our as well as in a previous
study~\cite{Gao:02}, two hydrogen bonds near the C terminus break
early in some cases, which reduces the number of FG bonds to
$\nfg=4$. The transition frequencies seen in Fig.~4 match well with
the ordering $\nbe\sim\nfg<\nab<\ncf$.  The first branch point in
Fig.~4 is the native state. Transitions from this state to the G
state, N$\to$G, are more common than N$\to$A transitions, in line with
the relation $\nfg<\nab$. The second layer of branch points is the A
and G states. That transitions G$\to$AG are more common than G$\to$GF
and that A$\to$AG and A$\to$AB have similar frequencies, match well
with the relations $\nab<\ncf$ and $\nfg\sim\nbe$,
respectively. Finally, there are fewer hydrogen bonds connecting the
AB hairpin to the rest of the native structure than what is the case
for the FG hairpin, $\nbe<\ncf$, which may explain why the AB hairpin,
unlike the FG hairpin, detaches as one unit in some runs.

Another feature seen from Fig.~4 is that the remaining native-like
core rotates during the course of the unfolding process.  The
orientation of the core is crucial, because a strand is much more
easily released if it can be unzipped one hydrogen bond at a time,
rather than by longitudinal pulling.  The detachment of the first
strand leads, irrespective of whether it is A or G, to an arrangement
such that two strands are favorably positioned for unzipping, which
explains why the intermediates with only A or G detached have a low
force-resistance (see Tables~3 and 4).  The AG, AB and FG
intermediates, on the other hand, have cores that are pulled
longitudinally, which makes them more resistant.  Also worth noting is
that the core of the AG intermediate is flipped 180$^\circ$, which is
not the case for the AB and FG intermediates.

The end-to-end distance of the intermediates cannot be directly
compared with experimental data. The experiments measured
~\cite{Li:05} contour-length differences rather than $L$, through
worm-like chain (WLC)~\cite{Marko:95} fits to constant-velocity data.
Using data at our lowest pulling velocity (0.03\,fm/MC step), we now
mimic this procedure. For each force peak, we determine a contour
length $\Lc$ by fitting the WLC expression
\begin{equation}
  F = \frac{\kb T}{\xi} 
  \left[ \frac{1}{4(1-z/\Lc)^2}-\frac{1}{4}+ \frac{z}{\Lc}
    \right]
\label{WLC}
\end{equation}
to data. Here $\xi$ denotes the persistence length and $z$ is the
elongation, defined as $z=L-\LN$, where $\LN$ is the end-to-end
distance of the native state.  Following \cite{Li:05}, we use a fixed
persistence length of $\xi=0.4$ nm.

After each rupture peak follows a region where the force is relatively
low. Here it sometimes happens that the newly released chain
segment forms $\alpha$-helical structures, indicating that our system
is not perfectly described by the simple WLC model.  Nevertheless, the
WLC model provides a quite good description of our unfolding traces,
as illustrated by Fig.~5. The figure shows a typical unfolding
trajectory with three force peaks, corresponding to the native (N),
intermediate (I) and unfolded (U) states, respectively. From the
fitted $\Lc$ values, the contour-length differences $\DLcNI$, $\DLcIU$
and $\DLcNU$ can be calculated.

\begin{figure}
  \begin{center}
    \includegraphics[width=0.47\textwidth,clip=true]{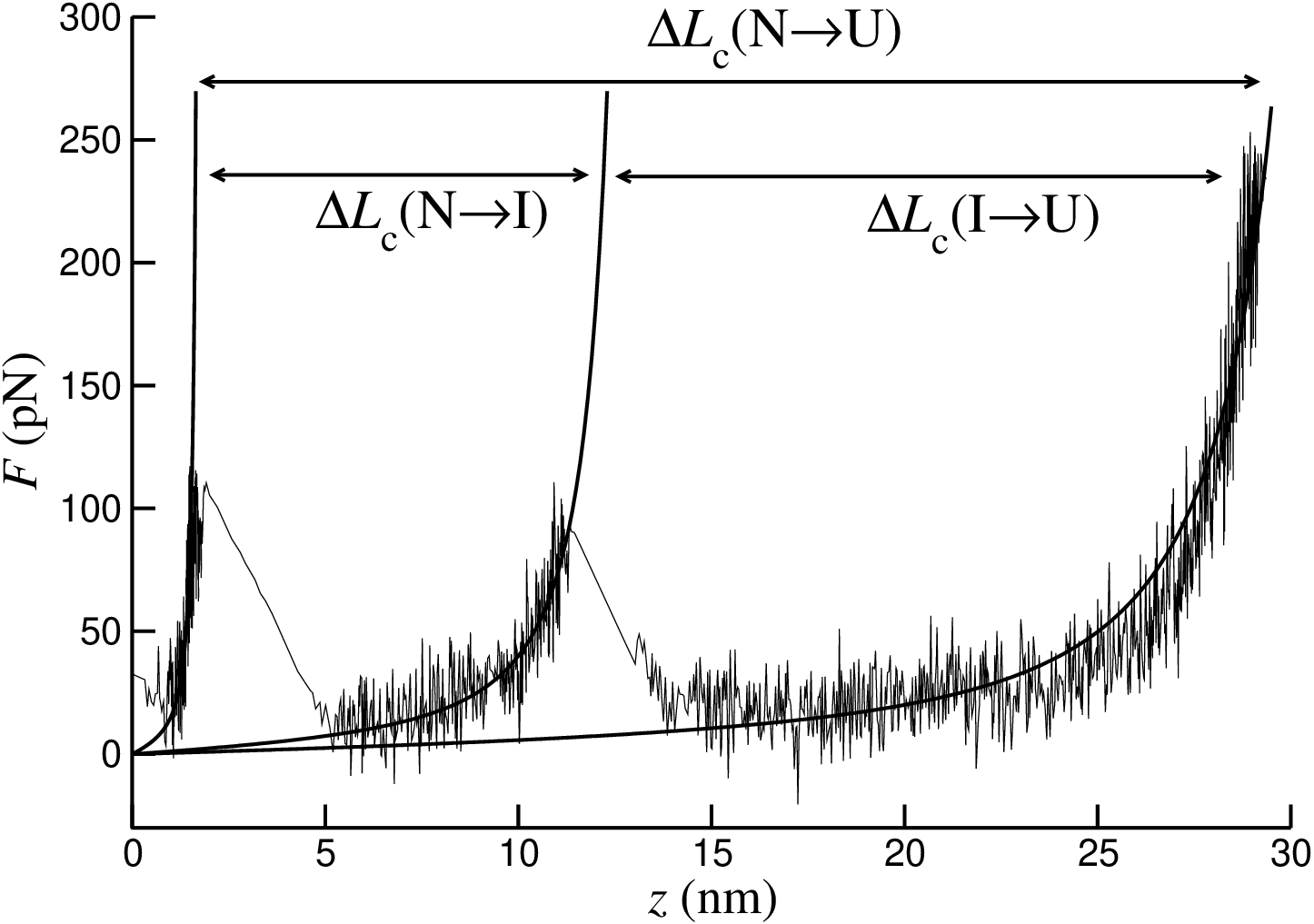}
  \end{center}
    \caption{WLC fits (Eq.~\ref{WLC}) to a typical force-extension
      curve at $v=0.03$\,fm/MC~step. The arrows indicate
      contour-length differences extracted from the fits: $\DLcNI$,
      $\DLcIU$ and $\DLcNU$.  }
\end{figure}

Fig.~6 shows a histogram of $\DLcNI$, based on our 100 trajectories
for $v=0.03$\,fm/MC step. For a small fraction of the force peaks, a
WLC fit is not possible; e.g., the A state cannot be analyzed due to
its closeness to the native state. All intermediates analyzed have a
$\DLc(\text{N}\to\text{I})$ in the range 6--27\,nm.  They are divided
into five groups: AB, AG, FG, G and ``other''.  Most of those in the
category ``other'' have five strands detached (CDEFG or ABEFG) and a
$\DLcNI$ larger than 21\,nm. These intermediates were not identified
in the experimental study~\cite{Li:05}, which did not report any
$\DLcNI$ values larger than 18\,nm.  These high-$L$ intermediates
mainly occur as a second intermediate, following one of the main
intermediates, which perhaps explains why they were not observed in
the experiments.  The few remaining intermediates in the category
``other'' are all of the same kind, ABG, but show a large variation in
$\DLcNI$, from 10 to 19\,nm.  The small values correspond to states
where strand B actually is attached to the structured core, but
through non-native hydrogen bonds.

\begin{figure}
  \begin{center}
    \includegraphics[width=0.47\textwidth,clip=true]{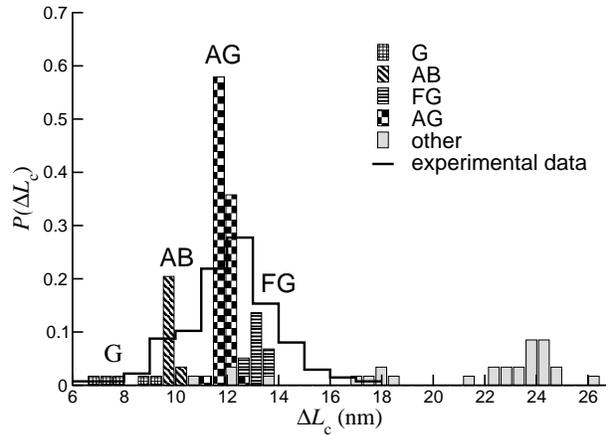}
  \end{center}
  \caption{Histogram of the contour-length difference $\DLcNI$,
    obtained by WLC fits (Eq.\ \ref{WLC}) to our data for
    $v=0.03$\,fm/MC~step. A total of 121 force peaks corresponding to
    intermediate states are analyzed. The intermediates are divided
    into five groups: AB, AG, FG, G and ``other''. The experimental
    $\DLcNI$ distribution, from \cite{Li:05}, is also indicated. }
\end{figure}

The three major peaks in the $\DLcNI$ histogram (Fig.~6) correspond to
the AG, AB and FG intermediates. Although similar in size, these
states give rise to well separated peaks, the means of which differ in
a statistically significant way (see Table~6). For comparison, Fig.~6
also shows the experimental $\DLc(\text{N}\to\text{I})$ distribution
~\cite{Li:05}. The statistical uncertainties appear to be larger in
the experiments, because the distribution has a single broad peak
extending from 6 to 18\,nm. All our $\DLcNI$ data for the AB, AG, FG
and G intermediates fall within this region. The occurrence of these
four intermediates is thus consistent with the experimental $\DLcNI$
distribution.  The highest peak, corresponding to the AG intermediate,
is located near the center of the experimental distribution.

\begin{table}
  \begin{center}
  \begin{footnotesize}
    \begin{tabular} {rc}
      \multicolumn{1}{r}{state}  &
      \multicolumn{1}{r}{$\avDLcNI$ (nm)}\\
      \hline\hline
         AG & $12.1\pm0.3$\\
         AB & $10.1\pm0.1$\\
         FG & $13.4\pm0.3$\\
         G  & $8.2\pm0.9$\\
      \hline\hline
    \end{tabular}
  \end{footnotesize}
  \end{center}
  \caption{The average contour-length difference $\avDLcNI$ for
    different intermediates, as obtained by WLC fits (Eq.~\ref{WLC})
    to our data for $v=0.03$\,fm/MC~step.}
\end{table}

Transitions from the native state directly to the unfolded state do
not occur in the trajectories analyzed for Fig.~6. For the
contour-length difference between these two states, we find a value of
$\DLcNU=30.9\pm 0.1$\,nm, in perfect agreement with experimental
data~\cite{Li:05}.

\subsection*{Estimating the free-energy profile}

We now present the free-energy profile obtained by applying
Eqs.~\ref{ej}--\ref{eqhis} to the constant-velocity trajectories. The
number of trajectories analyzed can be seen in Table~1.
Fig.~7 shows the free-energy landscape at zero force, $G_0(L)$, against the 
end-to-end distance $L$, as obtained using different velocities $v$. We 
observe a collapse of the curves in the region of small-to-moderate $L$. 
Furthermore, the range of $L$ where the curves superimpose, expands as $v$
decreases.  As discussed
in~\cite{Imparato:06,Imparato:07a,Imparato:07c,Imparato:07b}, the
collapse of the reconstructed free-energy curves, as the manipulation
rate is decreased, is a clear signature of the reliability of the
evaluated free-energy landscape. Given our computational resources, we
are not able to further decrease the velocity $v$, and for $L>15$ nm
there is still a difference of $\sim$\,40\,$\kb T$ between the two
curves corresponding to the lowest velocities.  The best estimate we
currently have for $G_0(L)$ is the curve obtained with 
$v=0.03$\,fm/MC~step. This curve will be used in the following analysis.

\begin{figure}
  \begin{center}
    \includegraphics[width=0.47\textwidth]{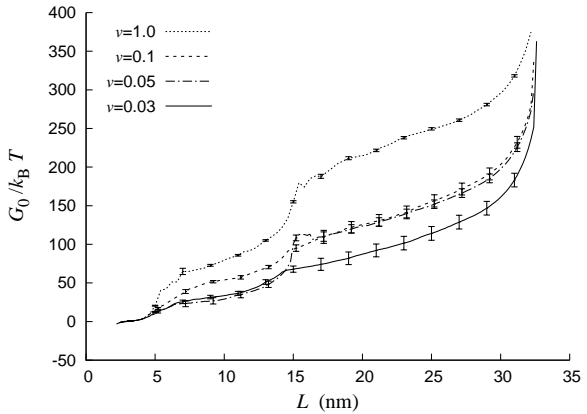}
  \end{center}  
  \caption{Free-energy landscape $G_0(L)$ calculated as a 
    function of the end-to-end distance $L$, using data at different 
    pulling velocities $v$ (given in fm/MC~step). In the calculations,
    $L$ is discretized with a bin size of $\Delta L=0.4$\,nm for 
    $v=1.0$\,fm/MC~step and $\Delta L=0.2$\,nm for all the other velocities
    (see Model and Methods).}
\end{figure}

Let us consider the case where a constant force $F$ is applied to the
chain ends.  The free energy then becomes $G(L) = G_0(L)-F\cdot L$.  
The tilted free-energy landscape $G(L)$ is especially interesting for 
small forces for which the unfolding process is too slow to be 
studied through direct simulation. 

Fig.~8 shows our calculated $G(L)$ for four external 
forces in the range 10--50\,pN. At $F=10$\,pN, the state with 
minimum free energy is still the native one, and no additional local 
minima have appeared. At $F=25$\,pN, the situation has
changed. For $20\lesssim F\lesssim 60$\,pN, we find that 
$G(L)$ exhibits three major minima: the native minimum and
two other minima, one of which corresponds to the fully unfolded state.   
The fully unfolded state takes over as the global minimum beyond 
$F=\Fc\approx22$\,pN. The statistical uncertainty on the force at which 
this happens, $\Fc$, is large, due to uncertainties on $G(L)$ for 
large $L$, as will be further discussed below. 
For $F=25$\,pN, the positions of the three 
major minima are $4.3$\,nm, 12\,nm and 25\,nm.  As $F$ increases, 
the minima move slightly toward larger $L$; for $F=50$\,pN, 
their positions are 4.6\,nm, 14\,nm and 29\,nm. 
The first two minima become increasingly shallow with increasing $F$.  
For $F\gtrsim60$\,pN, the only surviving minimum is the third one, 
corresponding to the completely unfolded state. 

\begin{figure}
  \begin{center}
    \includegraphics[width=0.47\textwidth]{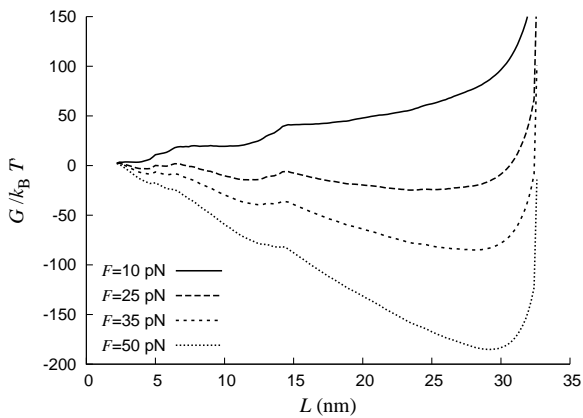}
  \end{center}  
  \caption{Tilted free-energy landscape $G(L)=G_0(L)-F\cdot L$
    for four different forces $F$. 
    The unperturbed landscape $G_0(L)$ corresponds to the curve shown 
    in Fig.~7 for $v=0.03$\,fm/MC~step. The minima of $G(L)$ are discussed 
    in the text.}
\end{figure}

These results have to be compared with the analysis above, which showed  
that the system, on its way from the native to the fully unfolded state, 
often spends a significant amount of time in some partially 
unfolded intermediate state with $L$ around 12--16\,nm. 
These intermediates should correspond to local 
free-energy minima along different unfolding pathways, but 
may or may not correspond to local minima of the global free energy 
$G(L)$, which is based on an average over the full conformational space. 
As we just saw, it turns out that $G(L)$ actually exhibits a
minimum around 12--16\,nm, where the most common intermediates 
are found. It is worth noting that above $\sim$\,25\,pN this minimum 
gets weaker with increasing force. This trend is in agreement
with the results shown in Table~2: the fraction of apparent two-state
events, without any detectable intermediate, increases with increasing
force.

For $F=25$\,pN and $F=35$\,pN, a fourth minimum can also be seen in
Fig.~8, close to the native state. Its position is $\approx$\,6\,nm.
This minimum is weak and has already disappeared for $F=50$\,pN. It
corresponds to a state in which the two native $\beta$-sheets are
slightly shifted relative to each other and aligned along the
direction of the force, with all strands essentially intact.  The
appearance of this minimum is in good agreement with the results of
Gao et al.~\cite{Gao:02}. In their unfolding trajectories, Gao et
al. saw two early plateaus with small $L$, which in terms of our
$G(L)$ should correspond to the native minimum and this $L\approx
6$\,nm minimum. In our model, the $L\approx6$\,nm minimum represents a
non-obligatory intermediate state; in many unfolding events,
especially at high force, the molecule does not pass this state.

Finally, Fig.~9 illustrates a more detailed analysis of the native
minimum of $G(L)$, for $20\,\text{pN}<F<60$\,pN. In this force range,
we find that the first barrier is always located at $L=5.0$\,nm,
whereas the position of the native minimum varies with force (see
inset of Fig.~9).  Hence, the distance between the native minimum and
the barrier, $\xu$, depends on the applied force, as
expected~\cite{Li:03,Hyeon:06,West:06b,Dudko:07}.  Fig.~9 shows the
force-dependence of the barrier height, $\Delta G(F)$.  The solid line
is a linear fit with slope $\xu = 0.4$\,nm, which describes the data
quite well in the force range 25--56\,pN.  At lower force, the
force-dependence is steeper; a linear fit to the data at low force
gives a slope of $\xu = 0.8$\,nm (dashed line). Using this latter fit
to extrapolate to zero force, we obtain a barrier estimate of $\Delta
G(0)\approx 5$\,kcal/mol. Due to the existence of the non-obligatory
$L\approx 6$\,nm intermediate, it is unclear how to relate this
one-dimensional free-energy barrier to unfolding
rates. Experimentally, barriers are indirectly probed, using unfolding
kinetics. For \fntt, experiments found a zero-force barrier of
22.2\,kcal/mol~\cite{Oberhauser:02}, using kinetics.  For the
unfolding length, an experimental value of $\xu=0.38$\,nm was
reported~\cite{Oberhauser:02}, based on data in the force range
50--115\,pN. Our result $\xu=0.4$\,nm obtained using the overlapping
force range 25--56\,pN, is in good agreement with this value.
\begin{figure}
  \begin{center}
    \includegraphics[width=0.47\textwidth,clip=true]{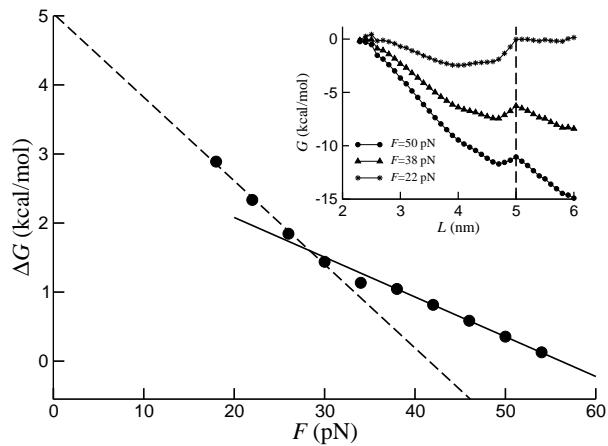}
  \end{center}
  \caption{Free-energy barrier $\Delta G$, separating the native state
    from extended conformations, as a function of the pulling force
    $F$. The solid line is a linear fit to the data for forces $ F >
    25$ pN, while the dashed line refers to a linear fit to the data
    in the interval $15\,\text{pN}\le F\le 30\,\text{pN}$.  The inset
    shows the free energy $G(L)$ in the vicinity of the native state
    for three values of the force. The vertical dashed line indicates
    the position of the barrier.}
\end{figure}

\section*{Discussion}

By AFM experiments, Li et al.~\cite{Li:05} showed that \fntt\ unfolds
through intermediates when stretched by an external force. AFM data
for the wild-type sequence and some engineered mutants were consistent
with the existence of two distinct unfolding pathways with different
intermediates, one being the AB state with strands A and B detached
and the other being either the AG or the FG state~\cite{Li:05}.  This
conclusion is in broad agreement with simulation results obtained by
Paci and Karplus~\cite{Paci:99} and by Gao et al.~\cite{Gao:02}.

Comparing our results with these previous simulations, one finds both
differences and similarities.  In our simulations, three major
intermediates are observed: AB, which was seen by Paci and Karplus as
well as by Gao et al.; AG, also seen by Paci and Karplus; and FG,
which was not observed in previous studies.  The most force-resistant
intermediate is AB in our as well as in previous studies. Frequencies
of occurrence of the intermediates are difficult to compare because
the previous studies were based on fewer trajectories. Nevertheless,
one may note that the most common intermediate in our simulations, AG,
is one of two intermediates seen by Paci and Karplus, and corresponds
to one of three pathways observed by Gao et al.  A and G often being
the first two strands to break is also in agreement with the
simulation results of Klimov and Thirumalai~\cite{Klimov:00}, who
studied several different proteins using a simplified model. Unlike
us, these authors found a definite unfolding order for the
$\beta$-strands. The first strand to break was G, followed by A.

A key issue in our study is how the unfolding pathway depends on the
pulling strength. This question was addressed by Gao et
al.~\cite{Gao:02}. Based on a simple analytical model rather than
simulations, it was argued that there is a single unfolding pathway at
low force and multiple unfolding pathways at high force.  Our results
show the opposite trend. At our lowest force, 50\,pN, we observe
several different unfolding pathways, and all the three major
intermediates occur with a significant frequency. At our highest
force, 192\,pN, unfolding occurs either in one step or through one
particular intermediate, the AG state. Moreover, at 192\,pN, the same
two strands, A and G, are almost always the first to break in the
apparent one-step events as well. Hence, at our highest force, we find
that the unfolding behavior has an essentially deterministic
component. The trend that the unfolding pathway becomes more
deterministic with increasing force can probably be attributed to a
reduced relative importance of random thermal fluctuations.

There is a point of disagreement between our results and experimental
data, which is that the rupture forces of the three major
intermediates are higher in our constant-velocity simulations than
they were in the experiments~\cite{Li:05}.  Although the statistical
uncertainties are non-negligible and the pulling conditions are not
identical (e.g., we consider a single \fntt\ module, while the
experiments studied multimodular constructs), we do not see any
plausible explanation of this discrepancy.  It thus seems that our
model overestimates the rupture force of these intermediates. Our
calculated rupture force for the native state is consistent with
experimental data (see above). To make sure that this agreement is not
accidental, we also measured the rupture force of the native state for
three other domains, namely \fntii, \fntiii\ and the titin I27
domain. AFM experiments (at 0.6\,$\mu$m/s) found that these domains
differ in force-resistance, following the order
\fntiii\,($\sim$\,90\,pN) $<$ \fntii\,($\sim$\,120\,pN) $<$
I27\,($\sim$\,200\,pN)~\cite{Oberhauser:02}.  For each of these
domains, we carried out a set of 60 unfolding simulations, at a
constant velocity of 0.10\,fm/MC step. The average rupture forces were
$108\pm 4$\,pN for \fntiii, $135\pm4$\,pN for \fntii, and
$159\pm6$\,pN for I27, which is in reasonable agreement with
experimental data.  In particular, our model correctly predicts that
the force-resistance of the native state decreases as follows: I27 $>$
\fntii $>$ \fntiii $\sim$ \fntt. Similar findings have been reported
for another model~\cite{Craig:04}.
 
Throughout the paper, times have been given in MC steps. In order to
roughly estimate what one MC step corresponds to in physical units, we
use the average unfolding time of the native state, which is
$\sim$\,$4\cdot10^8$\,MC steps at our lowest force, 50\,pN. Assuming
that the force-dependence of the unfolding rate is given by $k(F) =
k_0 \exp(F \xu/\kb T)$~\cite{Bell:78} with
$\xu=0.38$\,nm~\cite{Oberhauser:02}, this unfolding time corresponds
to a zero-force unfolding rate of $k_0\sim
1/(4\cdot10^{10}\,\textrm{MC steps})$.  Setting this quantity equal to
its experimental value,
$k_0=0.02\,\textrm{s}^{-1}$~\cite{Oberhauser:02}, gives the relation
that one MC step corresponds to $1\cdot10^{-9}$\,s. Using this
relation to translate our pulling velocities into physical units, one
finds, for example, that 0.05\,fm/MC step corresponds to
0.05$\mu$m/s. This estimate suggests that the effective pulling
velocities in our simulations are comparable to or lower than the
typical pulling velocity in the experiments~\cite{Li:05}, which was
$0.4\mu$m/s.  That the effective pulling velocity is low in our
simulations is supported by the observation made earlier that the
force drops to very small values between the rupture peaks.

The force range studied in our simulations is comparable to that
studied in AFM experiments~\cite{Oberhauser:02,Li:05,Ng:07}.  The
exact forces acting on fibronectin under physiological conditions are
not known, but might be considerably smaller. For comparison, it was
estimated that physiologically relevant forces for the muscle protein
titin are $\sim$\,4\,pN per I-band molecule~\cite{Li:02}.  For so
small forces, the unfolding of \fntt\ occurs too slowly in the model
to permit direct simulation. Therefore, we cannot characterize
unfolding pathways and possible intermediates for these forces.  On
the other hand, we have an estimate of the free-energy profile $G(L)$
for arbitrary force, which can be used, in particular, to estimate the
force $\Fc$, beyond which the fully extended state has minimum free
energy. Using our best estimate of $G(L)$, one finds an $\Fc$ of
22\,pN (see above). Now, $\Fc$ depends on the behavior of $G(L)$ for
large $L$, where the uncertainties are large and not easy to
accurately estimate.  As a test, we therefore repeated the same
analysis using the Ising-like model of
\cite{Imparato:07a,Imparato:07c,Imparato:08b} (unpublished results),
which gave us the estimate $\Fc\sim20$\,pN, in quite good agreement
with the value found above (22\,pN). Together, we take these results
to indicate that $\Fc\gtrsim15$\,pN, which might be large compared to
physiologically relevant forces (see above). For stretching forces $F$
significantly smaller than $\Fc$, the statistical weight of the fully
stretched state is small. To estimate the suppression, let $\LN$ and
$\Ls$ be the end-to-end distances of the native and stretched
states. The free energies of these states at force $F$ can be written
as $\GN=\GNc-(F-\Fc)\LN$ and $\Gs=\Gsc-(F-\Fc)\Ls$, where $\GNc$ and
$\Gsc$ are the free energies at $\Fc$. Assuming $\GNc=\Gsc$,
$\Fc\gtrsim15$\,pN and $\Ls-\LN\sim 20$\,nm, one finds, for example,
that $\Gs-\GN\gtrsim25\,\kb T$ for $F\lesssim10$\,pN.  Our estimate
$\Fc\gtrsim15$\,pN thus indicates that unfolding of \fntt\ to its
fully stretched state is a rare event for stretching forces $F\lesssim
10$\,pN. The major intermediates are also suppressed compared to the
native state for $F\lesssim 10$\,pN (see Fig.~8). However, our results
indicate that the major intermediates are more likely to be observed
than the fully stretched state for these forces.

By extrapolating from experimental data at zero force, the force at
which the native and fully stretched states have equal free energy has
been estimated to be 3.5--5\,pN for an average FnIII
domain~\cite{Erickson:94}. Our results suggest that the native state
remains thermodynamically dominant at so small forces.

The reconstructed free energies $G_0(L)$ and $G(L)$ are
thermodynamical potentials describing the equilibrium behavior of the
system in the absence and presence of an external force $F$,
respectively.  On the other hand, the long-lived intermediate states
observed during the unfolding of the molecule are a clear signature of
out-of-equilibrium behavior. They indicate an arrest of the unfolding
kinetics, typically in the $L$ range 12-16\,nm, on the way from the
old (native) equilibrium state to the new (fully unfolded) equilibrium
state. The calculated (equilibrium) landscape $G(L)$ (see Fig.~8) is
to some extent able to describe this out-of-equilibrium behavior.  For
$20\lesssim F\lesssim 60$\,pN, this function exhibits three major
minima corresponding to the folded state, the most common
intermediates, and the fully unfolded state, respectively.  However,
since $G(L)$ describes the system in terms of a single coordinate $L$
and ``hides'' the microscopic configuration, one cannot extract the
full details of individual unfolding pathways from this function. For
example, one cannot, based on $G(L)$, distinguish the AG, AB and FG
intermediates, which have quite similar $L$.

The height of the first free-energy barrier, $\Delta G$, can be
related to the unfolding length $\xu$, a parameter typically extracted
from unfolding kinetics, assuming the linear relationship $\Delta G(F)
= \Delta G_0-F \cdot \xu$.  The parameter $\xu$ measures the distance
between the native state and the free-energy barrier, which generally
depends on force. Our data for $\xu$ indeed show a clear
force-dependence (see inset of Fig.~9). However, over a quite large
force interval, our $\xu$ is almost constant and similar to its
experimental value~\cite{Li:05}, which was based on an overlapping
force interval.

\section*{Conclusion}

We have used all-atom MC simulations to study the force-induced
unfolding of the fibronectin module \fntt, and in particular how the
unfolding pathway depends on the pulling conditions.  Both at constant
force and at constant pulling velocity, the same three major
intermediates were seen, all with two native $\beta$-strands missing:
AG, AB or FG. Contour-length differences $\DLcNI$ for these states
were analyzed, through WLC fits to constant-velocity data. We found
that the states, in principle, can be distinguished based on their
$\DLcNI$ distributions, but the differences between the distributions
are small compared to the resolution of existing experimental data.

The unfolding behavior at constant force was examined in the range 
50--192\,pN. The following picture emerges from this analysis:

\begin{enumerate}
\item At the lowest forces studied, several different unfolding
pathways can be seen, and all the three major intermediates occur with
a significant frequency. \vspace{-6pt}

\item At the highest forces studied, the AB and FG intermediates are
very rare. Unfolding occurs either in an apparent single step or
through the AG intermediate.\vspace{-6pt}

\item The unfolding behavior becomes more deterministic with
increasing force.  At 192\,pN, the first strand pair to break is
almost always A and G, also in apparent two-state events.
\end{enumerate}   

The dependence on pulling velocity in the constant-veloc\-ity
simulations was found to be somewhat less pronounced, compared to the
force-dependence in the constant-force simulations. Nevertheless, some
clear trends could be seen in this case as well. In particular, with
increasing velocity, we found that the AG state becomes increasingly
dominant among the intermediates. Our results thus suggest that the AG
state is the most important intermediate both at high constant force
and at high constant velocity.

The response to weak pulling forces is expensive to simulate; our
calculations, based on a relatively simple and computationally
efficient model, extended down to 50\,pN. The Jarzynski method for
determining the free energy $G(L)$ opens up a possibility to partially
circumvent this problem. Our estimated $G(L)$, which matches well with
several direct observations from the simulations, indicates, in
particular, that stretching forces below 10\,pN only rarely unfold
\fntt\ to its fully extended state. Although supported by calculations
based on a different model, this conclusion should be verified by
further studies, because accurately determining $G(L)$ for large $L$
is a challenge.

\subsubsection*{Acknowledgments}
\vspace{-6pt} This work has been in part supported by the Swedish
Research Council and by the European Community via the STREP project
EMBIO NEST (contract no. 12835).
\end{small}
\begin{footnotesize}

\end{footnotesize}
\end{document}